\documentclass[showpacs,preprintnumbers,prd,nofootinbib,floats,amssymb,floatfix]{revtex4}
\usepackage{amsmath}
\usepackage{amsfonts}
\usepackage{graphicx}
\usepackage{hyperref}

\usepackage{amsmath}
\usepackage{amstext}
 
\begin{document}

\title{  A pure Dirac's canonical analysis   for four-dimensional  BF  theories    }  
\author{ Alberto Escalante }  \email{aescalan@sirio.ifuap.buap.mx}
 \affiliation{ 
 Instituto de F{\'i}sica Luis Rivera Terrazas, Benem\'erita Universidad Aut\'onoma de Puebla, (IFUAP).
   Apartado postal      J-48 72570 Puebla. Pue., M\'exico
    }
    \author{ I. Rubalcava-Garc{\'i}a}
    \affiliation{Facultad de Ciencias F\'{\i}sico Matem\'aticas, Universidad
Aut\'onoma de Puebla, Apartado postal 1152, 72001 Puebla Pue.,
M\'exico}. \email{}

\begin{abstract} 
 We perform Dirac's canonical analysis for a four-dimensional $BF$  and  for a  generalized four-dimensional $BF$ theory depending on   a connection valued in the Lie algebra of $SO(3,1)$.  This analysis is developed  by considering the corresponding  complete set of variables that  define these  theories   as dynamical, and we  find out the relevant symmetries,  the constraints,  the extended Hamiltonian, the extended action, gauge transformations  and the counting of physical degrees of freedom.  The results obtained are compared with other approaches  found in the literature.
\end{abstract}
\date{November 18,  2010}
\preprint{}
\maketitle
\section{INTRODUCTION}
\vspace{1em} \
Presently, the study of topological field theories is a topic of great interest in physics. The importance to study these theories arises  because they have a close relationship  with general relativity. These  theories  are  characterized by the absence  of local physical degrees of freedom and by the  background independence \cite{1}.  Relevant examples with  close  symmetries with general relativity are the so  called  $BF$ theories, which  are background independent and diffeomorphisms covariant, and  were introduced as generalizations of three dimensional Chern-Simons action or as a zero coupling limit of Yang-Mills theories \cite{2, 3}. We can find several examples where the $BF$ theories come to be physically relevant in  alternative formulations of gravity,  such as  Pleba\'nski  or  Macdowell-Mansouri formulations;  the former  consists in to obtain General Relativity  by imposing extra constraints on a $BF$ theory  with the gauge group $SO(3,1)$ or $SO(4)$ \cite{4}. The later  consists  in  breaking down the symmetry of a $BF$ theory from $SO(5)$ to $SO$(4),  to obtain  the Palatini action plus the sum of  the second Chern  and Euler topological invariants \cite{5}, and since  these topological classes  have trivial local variations  that  do not contribute classically to the dynamics, one  obtains  essentially general relativity  \cite{6}. \\
Other interesting case, where $BF$ theories have a close relation with physical theories is found in  Martellini's model \cite{7}.  This model consists in expressing  Yang-Mills theory as a $BF$-like theory, and the  BF first-order formulation is equivalent (on shell) to the usual (second-order) formulation. In fact, both formulations of the theory possess the same perturbative quantum properties; specifically  the Feynman rules, the structure of one loop divergent diagrams  and renormalization have been studied,  founding  an equivalence of the $uv$-behavior  for both approaches \cite{7}. Furthermore,  other kind of topological $BF$ theories are  reported  in \cite{8}, where by using a generalized differential calculus  a geometrical relation  between a generalized Chern-Simons functional  and a  generalized $BF$ theory is obtained; this  version  corresponds to a pure $BF$ term, plus the second Chern class  and a cosmological-like term quadratic in the field $B$. The case  of gravity viewed as a generalized topological field theory  and  its  close relation with the generalized $BF$ theory is also discussed \cite{8}.  \\
In this manner, with these motivations  we will perform the Hamiltonian analysis for a  four-dimensional $BF$ theory and for the generalized $BF$ theory introduced  in \cite{8}. It is important to remark that the standard way to develop the Hamiltonian study for a $BF$ theory is considering as dynamical variables only those ones that  occur in the Lagrangian density with temporal derivative \cite{9,10} (called smaller phase space context). However this approach is only convenient to perform provided that  the theory under study presents certain simplicity; but the price to pay  for developing  the standard approach is that we can not to know the full structure of the constraints and their  algebra, the equations of motion and the gauge transformations. Nevertheless, the  approach developed in this work will be quite different  to the standard one;  this means that   in agreement with the background independence structure that presents the theory under study,    we will develop the Hamiltonian framework by considering all the fields occurring in the theories  as dynamical ones;  this fact will allow us to find the complete structure of the constraints, the equations of motion, gauge transformations, the extended action as well as the extended Hamiltonian. We able to realize  that developing the Hamiltonian approach on a smaller phase space context,   the structure obtained for  the constraints is not right.  In fact,  we  observe in \cite{11b} that the Hamiltonian constraint for Palatini theory does not has the required structure  to   form a closed algebra with all  constraints;  this problem emerges because of  by working on a smaller phase space context  we  lose  control on the constraints,  and   to obtain  the correct structure sometimes they need to be fixed by hand  as it  was done in \cite{10} for Pleba\'nski theory. Nevertheless there are  analysis on a smaller phase space  performed without complications,  as for instance in Maxwell and  Yang-Mills theories \cite{11}.  For these reasons in this paper we develop  a pure Dirac method applied to models with a close relationship to Palatini and Pleba\'nski theory just as the four-dimensional BF theories, and we will see that  is not necessary to fix by hand the constraints because the method  itself provides  us the required structure. Thus in this paper we establish   the bases for   forthcoming works where  will be  applied  the same approach  to  Pleba\'nski theory.  We will discuss all these details along the paper and we have added an appendix to clarify these ideas. \\
 \newline
\newline
\setcounter{equation}{0} \label{c2}
\section{A pure Dirac's analysis   for  a four-dimensional BF theory}
In this section, we will develop  an extension  of the results reported in \cite{9}.  In the following lines, we shall study the Hamiltonian dynamics for a  four-dimensional $BF$ theory by using a pure Dirac's  method.  With the terminology "a pure Dirac's method" we mean  that in concordance with the background independence of the theory, we will consider in the Hamiltonian framework that all the fields that define our theory are  dynamical ones. \\
So, let us start with a four-dimensional $BF$ theory which is  described by the following  action \cite{9,10}
\begin{equation}
S \left[ A, 	\textbf{B} \right]= \int_M \textbf{B}^{IJ} \wedge \textbf{F}_{IJ} ( A ), 
\end{equation}
where  $F^{IJ}= d A_{IJ} + A{_{I}}^K \wedge A_{KJ}$  is the curvature of the Lorentz connection 1-form $A^{IJ}=A{_{\alpha}}^{IJ}dx^{\alpha}$,  and $B^{IJ}= \frac{1}{2}B{^{IJ}}_{\alpha \beta} dx^{\alpha}\wedge dx^{\beta}$ is a set of six $SO(3,1)$ valued 2-forms. Here, $\mu, \nu=0,1,..,3$ are spacetime indices, $x^\mu$  are the coordinates that label the points for the four-dimensional Minkowski manifold $M$ and  $I, J= 0,1..,3$ are internal indices that can be raised and lowered by the internal Lorentzian   metric  $\eta_{IJ}= (-1,1,1,1)$.\\
The equations of motion that arises from the variation of the action are given by 
\begin{eqnarray}
F&=&0, \nonumber \\
DB&=&0,  
\end{eqnarray}
in this sense, both $B$ and $A$ are considered as dynamical variables and we will take account this fact for all our developments along the paper. \\
To perform the Hamiltonian framework,  we will suppose that the manifold $M$ has topology $\Sigma \times R$, where $\Sigma$ corresponds to  Cauchy's surfaces and $R$ represents an evolution  parameter.   By performing  the 3+1 decomposition, we can write the action as 
\begin{equation}
S \left[ A, 	\textbf{B} \right]= \frac{1}{2}\int_R \int_\Sigma  dt d^3x \left\{ B{^{IJ}}_{0a} F_{IJ bc} \epsilon^{0abc} + 
\epsilon^{0abc}B_{IJab} \left( \dot{A}^{IJ}{_{c}}  -  \partial_c A^{IJ}{_{0}} - A^{IL}{_{0}} A_L{^{J}}{_{c}} + 
A^{IL}{_{c}} A_L{^{J}}{_{0}}   \right)  \right\},
\end{equation}
where  can be  identified   the following  Lagrangian density 
\begin{equation}
{\mathcal{L}}=  \frac{1}{2}B{^{IJ}}_{0a} F_{IJ bc} \epsilon^{0abc} +\frac{1}{2} \epsilon^{0abc}B_{IJab} \left( \dot{A}^{IJ}{_{c}}  -  
\partial_c A^{IJ}{_{0}} - A^{IL}{_{0}} A_L{^{J}}{_{c}} + A^{IL}{_{c}} A_L{^{J}}{_{0}} \right).
\label{23}
\end{equation}
As was commented earlier, the cornerstone of this work is to carry out  the Hamiltonian analysis  by  considering  as dynamical  all the set of  $A{_{\alpha}}^{IJ}\rightarrow$24  and  $B{^{IJ}}_{\alpha \beta}\rightarrow$36 variables  that define the action. We  observe at this point,  that our procedure is quite different to \cite{9,10} because in that work the Hamiltonian analysis was performed  considering  as  dynamical variables only those with time derivative occurring explicitly in the Lagrangian (in this particular scenario  only the  $A^{IJ}{_{c}}\rightarrow$18  are considered  dynamical).  It is important to observe also that strictly speaking,  $A^{IJ}{_{\alpha}}$ and $B^{IJ}{_{\alpha \beta}}$  are our set of dynamical variables and the correct form to carry out the Hamiltonian analysis is taking to account  that set. However, because of the action is not quadratic in the field $B_{0i}^{IJ}$ or $A_0^I$,  the Hamiltonian study has been performed on a smaller phase space context neglecting  the variables $A^{IJ}_{0}$ and  $B^{IJ}{_{0i}}$ as dynamical and being identified   as Lagrange multiplier \cite{9,10}. Nevertheless, it is not ever easy   perform the Hamiltonian analysis on a smaller phase space;  example of this fact is present in  Pleba\'nski's formulation \cite{10},   where Dirac's analysis needs a treatment with more details,  introducing new variables and fixing the structure of the first class constraints by hand. In this manner, we   attempt  with    the present  paper to extend the standard  approach developed for $BF$ theories,  by performing  a  full   canonical analysis  that will be useful in order  to carry out the analysis  for   Plebanski's  formulation or for models found in $BF$ gravity \cite{19}   without fixing  by hand the constraints as reported in \cite{10}. \\
 Hence, by identifying our set of dynamical variables,  a pure Dirac's method calls for the definition of the momenta 
 $\left(\Pi^\alpha {_{IJ}}, 
\Pi^{\alpha \beta}{_{IJ}} \right)$ canonically conjugate to $\left(A^{IJ}{_{\alpha}}, B{_{\alpha \beta}}^{IJ} \right)$, 
\begin{equation}
\Pi^\alpha {_{IJ}}= \frac{\delta L}{\delta \dot{A}^{IJ}{_{\alpha}} }, \quad  \Pi^{\alpha \beta}{_{IJ}}=\frac{\delta L}{\delta \dot{B}{_{\alpha \beta}}^{IJ} }.
\label{eq5}
\end{equation}
The matrix elements of the Hessian 
\begin{equation}
\frac{\partial^2{\mathcal{L}} }{\partial( \partial_\mu (A_{\alpha} {^{IJ}} )) \partial(\partial_\mu (A_{\beta} {^{IJ}} )) },\quad \quad \frac{\partial^2{\mathcal{L}} }{\partial( \partial_\mu (A_{\alpha} {^{IJ}} ) )\partial(\partial_\mu (B{_{\rho \nu}}^{IJ} ) ) },  \quad \quad \frac{\partial^2{\mathcal{L}} }{\partial( \partial_\mu (B{_{\rho \nu}}^{IJ} )) \partial(\partial_\mu (B{_{\gamma \sigma}}^{IJ}) ) }, 
\label{eq47}
\end{equation}
are identically zero, the rank of the Hessian is zero, thus, we expect 60 primary constraints. \\
From the definition of the momenta (\ref{eq5}),  we   identify the following  60 primary constraints
\begin{eqnarray}
\phi^0{_{IJ}} &:&   \Pi^0 {_{IJ}} \approx 0, \nonumber \\
\phi^a{_{IJ}} &:& \Pi^a {_{IJ}} - \frac{1}{2} \eta^{abc} B_{bc IJ} \approx 0,  \nonumber \\
\phi^{0a}{_{{IJ}}} &:& \Pi^{0 a}{_{IJ}} \approx 0, \nonumber \\
\phi^{ab}{_{{IJ}}} &:&\Pi^{ab}{_{IJ}} \approx  0, 
\label{eq6}
\end{eqnarray}
where we have defined  $\epsilon^{0abc} \equiv \eta^{abc}$.  \\
By neglecting terms  on the frontier, the canonical Hamiltonian for  $BF$ theory  is given by 
\begin{equation}
H_{c}= - \int  dx^3 \left[A^{IJ}{_{0}} D_a \Pi^a {_{IJ}} +\frac{1}{2}\eta^{abc}B{_{0a}}^{IJ} F_{IJ bc} \right] .
\label{eq7}
\end{equation}
In this manner,   adding  the primary constraints (\ref{eq6}) to the canonical Hamiltonian, the   primary Hamiltonian is given by 
\begin{equation}
H_P= H_c + \int dx^3 \left[  \lambda^{IJ}{_{0}} \phi{_{IJ}}^{0} + \lambda^{IJ}{_{a}} \phi{_{IJ}}^{a}+\lambda_{0a}{^{IJ}}\phi^{0a}{_{IJ}} +\lambda_{ab}{^{IJ}}\phi^{ab}{_{IJ}}  \right],
\label{eq8} 
\end{equation}
where $ \lambda^{IJ}{_{0}}$,  $\lambda^ {IJ}{_{a}}, \lambda_{0a}{^{IJ}}$ and $ \lambda_{ab}{^{IJ}}$  are Lagrange multipliers enforcing the constraints.\\
The non-vanishing fundamental Poisson brackets for the theory under study are given by 
\begin{eqnarray}
\{A{^{IJ}}_\mu(x^0, x), \Pi{^{\nu}}_{KL} (x^0, y) \} &=& \delta{^{\nu}}_{\mu} \frac{1}{2} \left( \delta{^{I}}_K \delta{^{J}}_L- \delta{^{I}}_L \delta{^{J}}_K  \right) \delta^3(x-y), \nonumber \\ 
\{B_{\alpha \beta}{^{IJ}}(x^0, x), \Pi{^{\mu \nu}}_{KL}(x^0,y) \} &=& \frac{1}{4} \left( \delta{^{\mu}}_{\alpha} \delta{^{\nu}}_\beta-\delta{^{\mu}}_\beta \delta{^{\nu}}_\alpha  \right) \left( \delta{^{I}}_{K} \delta{^{J}}_L -\delta{^{I}}_{L} \delta{^{J}}_K \right) \delta^3(x-y).
\label{eq9}
\end{eqnarray}
Now,  we need to identify whether  the  theory presents   secondary constraints. For this aim, we compute the  60 $\times$ 60 matrix whose entries are the Poisson brackets among the primary constraints (\ref{eq6}), the nonzero brackets are given by  
\begin{eqnarray}
\{ \phi^{a}{_{IJ}} (x), \phi^{bc} {_{KL}}(y) \}&=& -\frac{1}{4} \eta^{abc} (\eta_{IK}\eta_{JL}-\eta_{IL} \eta_{JK}   ) \delta^3(x,y),   
\label{eq10}
\end{eqnarray}
this matrix has rank= 36,  and 24  linearly independent null-vectors. This result suggests that  consistency   conditions  imply  24 secondary constraints.  From the temporal evolution of the constraints (\ref{eq6}) and the contraction with the 24 null vectors, it follows that  the following   24 secondary constraints arise 
\begin{eqnarray}
\dot{\phi}^{0}{_{IJ}}&=& \{\phi^{0}{_{IJ}} (x), {H}_{P} \} \approx 0 \quad \Rightarrow \quad \psi_{IJ}:= D_a \Pi^a{_{IJ}} \approx 0. \nonumber \\
\dot{\phi}^{0a}{_{IJ}}&= &\{\phi^{0a}{_{IJ}} (x), {H}_{P} \} \approx 0 \quad \Rightarrow \quad \psi{^{0a}}_{IJ}:= \frac{1}{2}\eta^{abc} F_{bcIJ }  \approx 0, 
\label{eq11}
\end{eqnarray}
and the   rank allows us fix the next    values for the Lagrange multipliers 
\begin{eqnarray}
\dot{\phi}^a{_{IJ}}= \{\phi^a{_{IJ}} (x), {H}_{P} \} \approx 0 \quad & \Rightarrow& \quad \left[\Pi^a{_{JL}}\eta_{KI}-\Pi^a{_{IL}}\eta_{KJ}  \right] A_0{^{KL}}+  \eta^{abc} D_b B_{0cIJ}  \nonumber \\ &-&\frac{1}{2} \eta^{abc} \lambda_{bcIJ}\approx 0, \nonumber \\
\dot{\phi}^{ab}{_{IJ}}= \{\phi^{ab}{_{IJ}} (x), {H}_{P} \} \approx 0 \quad &\Rightarrow& \quad  \eta^{abc} \lambda_{cIJ}  \approx 0. 
\label{eq12}
\end{eqnarray}
For this theory there are not third constraints. By following with the method, we need to separate from the primary and secondary constraints  which ones correspond to first and second class.  In order to achive  this aim, we need to calculate   the Poisson brackets  among   primary and secondary constraints, which the nonzero brackets are given by  
\begin{eqnarray}\label{eq13-1}
\{ \phi^{a}{_{IJ}} (x), \phi^{bc} {_{KL}}(y) \}&=& -\frac{1}{4} \eta^{abc} (\eta_{IK}\eta_{JL}-\eta_{IL} \eta_{JK}   ) \delta^3(x,y),  \nonumber \\    
\{ \phi^{a}{_{IJ}} (x), \Psi {_{KL}}(y) \}& =&\frac{1}{2}\left( \Pi^a{_{JL}} \eta_{IK} - \Pi^a {_{IL}} \eta_{KJ}  + \Pi^a {_{KJ}} \eta_{IL} - \Pi^a {_{KI}}\eta_{LJ}     \right) \delta^3(x-y), \nonumber \\
\{ \phi^{a}{_{IJ}} (x), \Psi^b {_{KL}}(y) \}& =&\frac{1}{2}\eta^{abc} \Big[ (\eta_{IK} \eta_{JL} - \eta_{KJ} \eta_{IL})\partial_c  + (A_{JLc} \eta_{IK} -A_{ILc} \eta_{KJ})\nonumber \\  & -& (A_{KI} \eta_{LJ} - A_{KJc} \eta_{LI} )\Big] \delta^3(x-y) , \nonumber \\
\{ \Psi{_{IJ}} (x), \Psi {_{KL}}(y) \}& =&- \frac{1}{2}( \Psi_{LJ} \eta _{IK} - \Psi_{KJ} \eta_{IL} + \Psi_{IL}\eta_{KJ} - \Psi_{IK}\eta_{LJ} )\delta^3(x-y) \approx 0, \nonumber \\
\{ \Psi{_{IJ}} (x), \Psi^a {_{KL}}(y) \}& =& \frac{1}{2}\left( \Psi^a{_{LJ}} \eta _{IK} - \Psi^a{_{KJ}} \eta_{IL} + \Psi^a{_{IL}}\eta_{KJ} -  \Psi^a{_{IK}}\eta_{LJ} \right) \delta^3(x-y).  \approx 0, 
\end{eqnarray}
Thus, we can observe that this matrix has a rank= 36 and 48   null-vectors.  In this manner, we find that our theory presents a set of 48 first class constraints and 36 second class constraints. By using the  contraction of the null vectors with the constraints (\ref{eq6}) and (\ref{eq11}), we identify the following  48 first class constraints 
\begin{eqnarray}
\gamma^0{_{IJ}} &:&   \Pi^0 {_{IJ}}, \nonumber \\
\gamma^{0a}_{{IJ}} &:& \Pi^{0 a}{_{IJ}},  \nonumber \\
\gamma_{IJ}&:& D_a \Pi^{a}{_{IJ}} - \eta_{abc} \left[ \Pi^a{_{IP}} \Pi^{bc}{_{QJ}}  \eta^{PQ} -   \Pi^a{_{JP}} \Pi^{bc}{_{QI}}  \eta^{PQ}  \right] ,  \nonumber \\
\gamma^a_{IJ} &:&  \frac{1}{2} \eta^{abc} F_{IJ bc} +2 D_c\Pi^{ca}{_{IJ}},
\label{eq14}
\end{eqnarray}
we   identify  the third  constraint as the Gauss constraint for $BF$ theory, corresponding  to the generator of $SO(3,1)$ transformations. It is important to remark,  that the null vectors obtained from (\ref{eq13-1})  provide   us the complete form  of the  first class constraints, and we have not fixed by hand their   structure.\\
The rank obtained form the matrix (\ref{eq13-1})  yields  identifying  the following  36 second class constraints
\begin{eqnarray}
\chi^a{_{IJ}} &:& \Pi^a {_{IJ}} - \frac{1}{2} \eta^{abc} B_{IJ bc},  \nonumber \\
\chi^{ab}{_{{IJ}}} &:&\Pi^{ab}{_{IJ}}. 
\label{eq15}
\end{eqnarray}
On the other hand,  we can observe that the 48 first class constraints  given in (\ref{eq14}) are not all independent. The reason for that is because of in virtue of Bianchi's  identity $DF_{IJ}=0$  one  finds that 
\begin{equation}
D_{a}\gamma^{a}_{{IJ}}-  \frac{1}{2}\left[ \Pi{^{ab}}_{IK} F{_{abJ}}^K -  \Pi{^{ab}}_{KJ} F{_{abI}}^K \right]=0.
\label{eq16}
\end{equation}
Thus, from the  $\gamma^{a}_{IJ}$=18  first class constraints   we identify that  [18-6]= 12  are  independent. Therefore,  we procede to calculate the physical degrees of freedom as follows; there are  120 canonical   variables, [48-6]=42  independent first class constraints   and 36 independent second class constraints. With this information,  we conclude that four-dimensional $BF$ theory is devoid  of local degrees of freedom, hence  in this sense we can say that the theory is topological; although this theory  has  global degrees of freedom due to the nontrivial topology of the manifold on which is defined \cite{17}. \\ 
Additionally we  observe that the complete structure of the constraints (\ref{eq14}) and (\ref{eq15}) as well as the full structure of reducibility conditions  were  not reported in \cite{9}.   The reason   is because in \cite{9,10}  Dirac's canonical method was performed on a smaller phase space context, thus the complete structure of the constraints and reducibility conditions were  not found. Of course, in our results, by considering   the second class constraints (\ref{eq15}) as strong equations the above results are reduced  to those  reported in \cite{9}, thus   our results  extend and complete those ones.  The correct identification of the constraints is a very important step because are used  to  carry out the counting of the physical degrees of freedom and  to identify the gauge transformations if there exist  first class constraints. On the other hand, the  constraints are  the guideline     to make the best progress for  the quantization of the theory. We need to remember that the   quantization  scheme   for  gauge theories as Maxwell or Yang-Mills can not be directly applied to  theories with the symmetry  of covariance  under diffeomorphisms  (as for  instance $ BF$ theories)  because we lose    relevant physical information \cite{11c}.\\
Now, we will calculate the algebra of  the constraints; smearing the constraints with test fields 
\begin{eqnarray}
\phi_1 &:=& \gamma^0{_{IJ}} \left[ {\mathbf{A}} \right]= \int dx^3  A^{IJ} \left[ \Pi^0{_{IJ}} \right], \nonumber \\
\phi_2  &:=&\gamma^{0a}{_{IJ}} \left[ {\mathbf{B}} \right]= \int dx^3 B_{a}{^{IJ}} \left[  \Pi^{0a}{_{IJ}}  \right], \nonumber \\
\phi_3  &:=& \gamma{_{IJ}} \left[ {\mathbf{C}} \right]= \int dx^3 C_a{^{IJ}}  \left[ D_a \Pi^{a}{_{IJ}} - \eta_{abc} \left[ \Pi^a{_{IP}} \Pi^{bc}{_{QJ}}  \eta^{PQ} -   \Pi^a{_{JP}} \Pi^{bc}{_{QI}}  \eta^{PQ}  \right] \right], \nonumber \\
\phi_4 &:=& \gamma^{0a}{_{IJ}} \left[ \mathbf{D} \right]= \int dx^3  {\mathbf{D}}{_{0a}}^{IJ} \left[ \frac{1}{2} \eta^{abc}F_{bcIJ}  - 2 D_b\Pi^{ab}{_{IJ}} \right], \nonumber \\
\phi_5  &:=&\chi^a{_{IJ}}  \left[ {\mathbf{H}} \right]= \int dx^3 H_a{^{IJ}} \left[ \Pi^{a}{_{IJ}} - \frac{1}{2}  \eta^{abc}B{_{bc}} ^{IJ} \right], \nonumber \\
\phi_6  &:=&\chi^{ab}{_{IJ}} \left[ {\mathbf{G}} \right]= \int dx^3 G{_{ab}}{^{IJ}} \left[ \Pi^{ab} {_{IJ}}  \right], 
\label{eq17}
\end{eqnarray} 
 the nonzero brackets  of   the constraints are given  by 
\begin{eqnarray}
\Big\{\phi_3\left[ {\mathbf{C}}{^{IJ}} \right],\phi_3\left[ {\mathbf{C}}'{^{KL}} \right]  \Big\} &=& \int dx^3\left[ {\mathbf{C}}^{IK} {\mathbf{C}}'_K{^{J}}-{\mathbf{C}}^{JK} {\mathbf{C}}'_K{^{I}} \right]  \gamma{_{IJ}} \approx0, \nonumber \\
\Big\{\phi_3\left[ {\mathbf{C}}{^{IJ}} \right],\phi_4\left[ {\mathbf{D}}{_{0a}}^{KL} \right]   \Big\} &=& \int dx^3\left[ {\mathbf{C}}^{IK}{\mathbf{D}}_{0aK}{^{J}}-{\mathbf{C}}^{JK}{\mathbf{D}}_{0aK}{^{I}} \right]  \gamma^{0a}{_{IJ}} \approx0, \nonumber \\
\Big\{\phi_3\left[ {\mathbf{C}}{^{IJ}} \right],\phi_5\left[ {\mathbf{H}}_a{^{KL}} \right]  \Big\} &=& \int dx^3\left[ {\mathbf{C}}^{IK} {\mathbf{H}}_{aK}{^{J}}- {\mathbf{C}}^{JK} {\mathbf{H}}{_{aK}}{^{I}} \right] \chi^a{_{IJ}}  \approx 0, \nonumber \\
\Big\{\phi_5 \left[ {\mathbf{H}}_{a}{^{IJ}}  \right] , \phi_6 \left[ {\mathbf{G}}'_{ab}{^{KL}}  \right]  \Big\}&=& -\frac{1}{2}\eta^{abc}\int dx^3 \left[ {\mathbf{H}}_{aKH} {\mathbf{G}}{_{bc}}^{KH}\right], 
\label{eq18}
\end{eqnarray}
where we are able to  appreciate that the constraints form a set of first and second class constraints as  expected. From  the constraint algebra (\ref{eq18}), we are able to   identify the Dirac brackets for the theory,  by   observing   that  the  matrix whose elements are only the Poisson brackets among   second class constraints  is given by 
\begin{equation}
C_{\alpha\beta} = \left(
\begin{array}{rr}
0 \qquad \qquad& -\frac{1}{4} \eta^{abc} (\eta_{IK}\eta_{JL}-\eta_{IL} \eta_{JK}   ) \delta^3(x-y)  \\
 \frac{1}{4} \eta^{abc} (\eta_{IK}\eta_{JL}-\eta_{IL} \eta_{JK}) \delta^3(x-y)   & 0 \qquad \qquad \\
 \label{eqa}
\end{array}
\right).
\end{equation}
In this manner,  the Dirac bracket among    two functionals $A$, $B$  is expressed   by 
\begin{equation}
\{A(x),B(y) \}_D= \{A(x),B(y)\}_P + \int du dv \{A(x), \zeta^\alpha(u) \} C^{-1}_{\alpha \beta}(u,v) \{\zeta^\beta(v), B(y) \}, 
\label{eq27}
\end{equation}
where $ \{A(x),B(y)\}_P$ is the usual Poisson bracket between the functionals $A,B$,   $\zeta^\alpha(u)=(\chi{_{IJ}}^{a}, \chi{_{IJ}}^{ab} ) $,  with   $C^{-1}_{\alpha \beta}(u,v)$  being   the inverse of (\ref{eqa}) which has a  trivial form. It is well known that Dirac's  bracket (\ref{eq27}) will be an essential ingredient   to make progress in the quantization of the theory \cite{13, 14}.\\
 Furthermore,  the identification of the constraints will allow us  to  identify the extended action. By using the first class constraints (\ref{eq14}),  the second class constraints (\ref{eq15}), and the Lagrange multipliers (\ref{eq12})  we find that the extended action takes the form
 \begin{eqnarray}
&S_E&\left[ A_{\alpha} {^{IJ}},\Pi^{\alpha}{_{IJ}}, B{_{\mu \nu}}^{IJ}, \Pi^{\mu \nu}{_{IJ}}, u_0{^{IJ}}, u_{0a}{^{IJ}}, u{^{IJ}}, u{_{a}}^{IJ}, v_a{^{IJ}}, v{_{ab}}{^{IJ}}  \right] =\int \Big\{ \dot{A}_{\alpha} {^{IJ}}\Pi^{\alpha}{_{IJ}}+ \dot{ B}{_{0a}}^{IJ}\Pi^{0a}{_{IJ}}\nonumber \\ 
&+& \dot{ B}{_{ab}}^{IJ}\Pi^{ab}{_{IJ}} -  H-u_0{^{IJ}} \gamma^0{_{IJ}}- u_{0a}{^{IJ}} \gamma^{0a}{_{IJ}} -    u{^{IJ}} \gamma{_{IJ}}-     u{_{a}}^{IJ}\gamma^{0a}{_{IJ}}    -v_a{^{IJ}} \chi^a{_{IJ}}     -v{_{ab}}{^{IJ}} \chi^{ab}{_{IJ}}         \Big\} dx^4, \nonumber \\
\label{eq19}
\end{eqnarray}
where  $H$ is   linear  combination of  first class constraints 
\begin{equation}
H= \frac{1}{2}A_0{^{IJ}}\left[ D_a \Pi^{a}{_{IJ}} - \eta_{abc} \left[ \Pi^a{_{IP}} \Pi^{bc}{_{QJ}}  \eta^{PQ} -   \Pi^a{_{JP}} \Pi^{bc}{_{QI}}  \eta^{PQ}  \right] \right]- B_{0a}{^{IJ}}\left[\frac{1}{2} \eta^{abc}F_{bcIJ}  - 2 D_b\Pi^{ab}{_{IJ}} \right],
\label{eq20}
\end{equation}
and $u_0{^{IJ}}, u_{0a}{^{IJ}}, u{^{IJ}}, u{_{a}}^{IJ}, v_a{^{IJ}}, v{_{ab}}{^{IJ}}$ are the Lagrange multipliers enforcing the first and second class  constraints. We can observe  that   by considering the second class constraints as strong equations   the Hamiltonian (\ref{eq20}) is reduced to the  Hamiltonian found in  \cite{9}  where was performed the Hamiltonian analysis on a smaller phase space context. In this manner,  we have   developed in this work    a best and complete  description at classical level.   \\
From the extended action we can identify the extended Hamiltonian given by
\begin{equation}
H_E= H-u_0{^{IJ}} \gamma^0{_{IJ}}- u_{0a}{^{IJ}} \gamma^{0a}{_{IJ}} - u{^{IJ}} \gamma{_{IJ}}- u{_{a}}^{IJ}\gamma^{0a}{_{IJ}}.
\end{equation}
It is well know  that the equations of motion obtained by means of the extended Hamiltonian, in general, are mathematically different from the Euler-Lagrange equations, but the difference is unphysical \cite{11}.\\
It is important to remark,  that  the theory under study has an  extended Hamiltonian which is  linear combination of first class constraints reflecting the general covariance of the theory,  just as General Relativity, thus, it is not possible to construct the Schrodinger equation because  the action of the Hamiltonian
on physical states is annihilation. In Dirac's quantization of systems with general covariance,  the restriction of our
physical state is archived by demanding that  the first class constraints in their  quantum form must be satisfied,  then  we can use  the tools   of Loop Quantum Gravity  by finding a set of  quantum states for the theory as was performed in  \cite{18} using the spin foam models. \\
We will continue  this section by computing the equations of motion obtained  from the extended action, which are expressed by 
\begin{eqnarray}
\delta A{_{0}}^{IJ}:  \dot{\Pi}^0{_{IJ}}&=&-\frac{1}{2}\left[ D_a\Pi^a{_{IJ}} -\eta_{abc} \left[ \Pi^a{_{IP}} \Pi^{bc}{_{QJ}}  \eta^{PQ} -   \Pi^a{_{JP}} \Pi^{bc}{_{QI}}  \eta^{PQ}  \right] \right] ,  \nonumber \\
\delta \Pi^0{_{IJ}}:  \dot{A}{_{0}}^{IJ} &=& u{_{0}}^{IJ},  \nonumber \\
\delta A{_{a}}^{IJ}: \dot{\Pi}^a{_{IJ}}&=&\left[A_{0J}{^{F}} + u_J{^{F}} \right] \Pi^a{_{IF}}-\left[A_{0I}{^{F}} + u_I{^{F}} \right] \Pi^a{_{JF}}+ \eta^{abc}\left[D_bB_{0cIJ}-D_bu_{cIJ} \right]  \nonumber \\ &+& 2\left[u_{bI}{^{F}}- B_{0bI}{^{F}}  \right]\Pi^{ab}{_{JF}}-2\left[u_{bJ}{^{F}}- B_{0bJ}{^{F}}  \right]\Pi^{ab}{_{IF}} , \nonumber \\
\delta \Pi^a{_{IJ}}:   \dot{A}{_{a}}^{IJ}&=&-D_a\left(\frac{1}{2}A_0{^{IJ}}+u^{IJ}  \right) + \left(u_a{^{IJ}} -B_{0a}{^{IJ}} \right)+ \frac{1}{2}\eta_{abc}\left[A_0{^{IL}}\Pi^{bc}{_{QL}}\eta^{JQ}- A_{0}{^{JL}}\Pi^{bc}{_{QL}}\eta^{IQ} \right] \nonumber \\
 &+& v_a{^{IJ}}, \nonumber \\ 
\delta B{_{0a}}^{IJ}:  \dot{\Pi}^{0a}{_{IJ}}&=& - \left[ \frac{1}{2} \eta^{abc}F_{bcIJ}  - 2 D_b\Pi^{ab}{_{IJ}} \right] ,  \nonumber \\
\delta \Pi^{0a}{_{IJ}}:  \dot{B}{_{0a}}^{IJ} &=&  u{_{0a}}^{IJ},  \nonumber \\
\delta B{_{ab}}^{IJ} : \dot{\Pi}^{ab}{_{IJ}}&=&- \frac{1}{2}\eta ^{abc}v_{cIJ}, \nonumber \\
\delta \Pi^{ab}{_{IJ}}:  \dot{B}{_{ab}}^{IJ}&=&   D_a\left( u_b{^{IJ}}-B_{0b}{^{IJ}} \right) - D_b\left( u_a{^{IJ}}-B_{0a}{^{IJ}} \right)+ v_{ab}{^{IJ}}, \nonumber\\  &-& \frac{1}{2} \eta_{gab}\left[\Pi^g{_{LK}}A_{0}{^{LJ}}\eta^{PI}-  \Pi^g{_{LK}}A_{0}{^{LI}}\eta^{PJ}   \right],  \nonumber \\
\delta u_0{^{IJ}} : \gamma^0{_{IJ}}&=&0, \nonumber \\
\delta  u_{0a}{^{IJ}}: \gamma^{0a}{_{IJ}}&=&0, \nonumber \\
\delta  u{^{IJ}}: \gamma{_{IJ}}&=&0, \nonumber \\
\delta u{_{a}}^{IJ}:\gamma^{0a}{_{IJ}} &=&0, \nonumber \\
\delta v_a{^{IJ}} :\chi^a{_{IJ}} &=&0, \nonumber \\
\delta v{_{ab}}{^{IJ}}: \chi^{ab}{_{IJ}}                                              &=&0 . 
\label{eq61}
\end{eqnarray} 
\newline
\newline
 \noindent \textbf{II. Gauge generator}\\[1ex]
 By following with our analysis, we need to know the gauge transformations on the phase space of the theory under study. For this important step, we shall use  Castellani's formalism which allows us to define  the following gauge generator in terms of the first class constraints (\ref{eq14})
 \begin{eqnarray}
G= \int_\Sigma \left[D_0 \varepsilon_0{^{IJ}} \gamma^0{_{IJ}}  + D_0 \varepsilon_{0a}{^{IJ}} \gamma^{0a}{_{IJ}} +\varepsilon^{IJ}\gamma{_{IJ}} + \varepsilon_a {^{IJ}}\gamma^{0a}{_{IJ}}  \right]dx^3, 
\label{eq62}
\end{eqnarray}
thus, we find that the   gauge transformations on the phase  space  are 
\begin{eqnarray}
\delta_0 A{_{0}}^{IJ} &=&D_0  \varepsilon{_{0}}^{IJ},  \nonumber \\
\delta_0 A{_{a}}^{IJ} &=& -  D_a  \varepsilon^{IJ},  \nonumber \\
\delta_0 B{_{0a}}^{IJ} &=&   D_0  \varepsilon{_{0a}}^{IJ},  \nonumber \\
\delta_0 B{_{ab}}^{IJ} &=& \left[ D_a \varepsilon{_{b}}^{IJ} - D_b\varepsilon{_{a}}^{IJ} \right] + \left[ \varepsilon^{IF} B_{abF}{^{J}}- \varepsilon^{JF}B_{abF}{^{I}} \right],  \nonumber \\
\delta_0 \Pi^0{_{IJ}}   &=&   0,  \nonumber \\
\delta_0 \Pi^a{_{IJ}}   &=& \left[\Pi^a{_{IL}}\varepsilon_J{^{L}}- \Pi^a{_{JL}}\varepsilon_I{^{L}}  \right]+ \eta^{adc}D_d\varepsilon_{cIJ} + 2\left[ \Pi^{ab}{_{KI}}\varepsilon_b {^{L}}_J- \Pi^{ab}{_{KJ}}\varepsilon_b {^{L}}_I \right],  \nonumber \\
\delta_0 \Pi^{0a}{_{I}}  &=&   0,  \nonumber \\
\delta_0 \Pi^{ab}{_{IJ}}  &=& - \left[\Pi^{ab}{_{IF}} \varepsilon^{F}{_{J}}- \Pi^{ab}{_{JF}} \varepsilon^{F}{_{I}} \right].  
\label{eq63}
\end{eqnarray}
We can see that, the diffeomorphisms are not present in the previous gauge transformations; however it is well known  that $BF$ theory is diffeomorphism covariant. Thus, the next question that arises is: how  can we  recover the diffeomorphisms symmetry from the above  gauge transformations?. The answer for this question can be found  redefining  the gauge parameters as $-\varepsilon{_{0}}^{IJ}=\varepsilon^{IJ}=- \xi^\rho A_\rho{^{IJ}}$ and $\varepsilon_{\mu}^{IJ}=- \xi^\rho B_{\mu \rho}{^{IJ}}$. In this manner  the gauge transformations (\ref{eq63}) take the following form 
\begin{eqnarray}
A'_{\mu}{^{IJ}} &\rightarrow& A_{\mu}{^{IJ}} + {\mathcal{L}}_\xi  A_{\mu}{^{IJ}} + \xi^{\rho}F_{\mu \rho}{^{IJ}} , \nonumber \\
B'_{\mu \nu}{^{IJ}}& \rightarrow& B_{\mu \nu}{^{IJ}} + {\mathcal{L}}_\xi  B_{\mu \nu}{^{IJ}} + \xi^{\rho} \left[ D_\nu B_{\mu \rho}{^{IJ}} +D_\mu B_{\rho \nu}{^{IJ}}+  D_\rho B_{\nu \mu }{^{IJ}} \right], 
\label{eq25}
\end{eqnarray}
which correspond to diffeomorphisms. Therefore, the latter correspond to an  internal symmetry of the theory. It is important to remark  that all this information has not been reported in the literature  with the details that we have developed, on the contrary, usually the people prefer to work on a smaller phase space context. Nevertheless, the price to pay for working on a smaller phase space is that we could not know all  the relevant information of the theory,   as the complete form of the constraints,  and  the complete form of the gauge transformations. In any case,  given a   theory whose symmetries we  wish study  one must;  first,   to develop  a pure Dirac's method,  and  then with all the information at hand,   we will able to  reproduce those  results that are obtained   under the smaller phase context.  All these ideas are being already applied to Pleba\'nski theories (see appendix), however the analysis is somewhat complicated  but the right structure of the constraints will be reported in forthcoming works \cite{13}.  \\
In the later section,  we shall develop  the Hamiltonian framework for a theory $BF$-like theory  that emerges from a generalized Chern-Simons theory.  
\newline
\newline
\newline
 \section{ A pure Dirac's method for a generalized  four-dimensional BF theory   }
 For this section,   the action under consideration  is given by 
\begin{equation}
S \left[ A, 	\textbf{B} \right]= \int_M \left[ \textbf{F}^{IJ}(A) \wedge \textbf{F }_{IJ}(A)+ 2k \textbf{B}^{IJ} \wedge \textbf{F}_{IJ} ( A ) +k^2 \textbf{B}^{IJ} \wedge \textbf{B}_{IJ} \right].    
\label{eq26}
\end{equation}
As commented in the introduction, the above action was obtained by using a generalized differential calculus, taking the generalized exterior derivative to a generalized Chern-Simons  form \cite{8}.  We are able to  observe that the first term in (\ref{eq26}) corresponds to second Chern-class, the second one  is a pure $BF$ term and the third one is identified as a cosmological-like  term \cite{9},  where  $k$ is a constant  \cite{8}. The Hamiltonian study for the action (\ref{eq26}) has not been reported in the literature. In this manner, the action (\ref{eq26}) is a good example for   applying  a pure Dirac's  method just as in above section. With the present analysis, we will be able to identify  the full symmetries of the theory;  we could think, however,  that the action (\ref{eq26}) being the  coupling  of topological terms, then   the complete  theory is topological as well. Nevertheless, the answer is not trivial because in the literature  we can find  examples where the coupling  of topological theories   is not topological any more,  since  there exist  physical degrees of freedom \cite{12}. Therefore we need to perform the Hamiltonian analysis to know the  symmetries of the theory.  For this  aim,  we will proceed    just as  in above section  developing  a pure  Hamiltonian framework. \\ 
By  taking  the variation of (\ref{eq26}) respect to our set of  dynamical variables $(A, B)$,    the equations of motion  are given by 
\begin{eqnarray}
D(F+kB)&=&0, \nonumber \\
k(F+ kB)=0, 
\label{eq27a}
\end{eqnarray}
from  Bianchi's identities $DF=0$ we can see that the second  equation implies the first one.  Thus, with that  learned in earlier sections  we expect for this theory reducibility conditions  among  the constraints,  just as in $BF$ theory. \\
By performing the 3+1 decomposition for all the terms of the action (\ref{eq26}) we obtain 
\begin{eqnarray}
\textbf{F}^{IJ} (A) \wedge \textbf{F }_{IJ}(A)&=& \frac{1}{4} \epsilon^{\alpha \beta \mu \nu} F^{IJ}{_{\alpha \beta}}F_{IJ \mu \nu} dx^4= \eta^{abc}F{_{bc}}^{IJ} \left( \dot{A}_{aIJ} - D_a A_{0IJ} \right), \nonumber \\
\textbf{B}^{IJ} \wedge \textbf{F}_{IJ} ( A ) &=& \frac{1}{4} \epsilon^{\alpha \beta \mu \nu} B^{IJ}{_{\alpha \beta}}F_{IJ \mu \nu} dx^4= \frac{1}{2}\eta^{abc}B{_{0a}}^{IJ}F_{bcIJ}+ \frac{1}{2}\eta^{abc}B{ _{bc}}^{IJ} \left( \dot{A}_{aIJ} - D_a A_{0IJ} \right), \nonumber \\
\textbf{B}^{IJ} \wedge \textbf{B}_{IJ} &=& \frac{1}{4} \epsilon^{\alpha \beta \mu \nu} B^{IJ}{_{\alpha \beta}}B_{IJ \mu \nu} dx^4= \eta^{abc}B{_{0a}}^{IJ}B{_{bc}}^{IJ}. 
\end{eqnarray}
In this way, the action principle takes the following  form 
\begin{eqnarray}
S \left[ A, 	\textbf{B} \right] &=&  \int_R \int_\Sigma [ \eta^{abc}F{_{bc}}^{IJ} \left( \dot{A}_{aIJ} - D_a A_{0IJ} \right)  +k \eta^{abc}B{_{0a}}^{IJ}F_{bcIJ}+ k \eta^{abc}B{ _{bc}}^{IJ} \left( \dot{A}_{aIJ} - D_a A_{0IJ} \right)  \nonumber  \\  &+&  k^2\eta^{abc}B{_{0a}}^{IJ}B{_{bc}}^{IJ} ] dx^3dt,    
\label{eq29a}
\end{eqnarray}
and the Lagrangian density is given by 
\begin{eqnarray}
{\mathcal{L}} &=& \eta^{abc}F{_{bc}}^{IJ} \left( \dot{A}_{aIJ} - D_a A_{0IJ} \right)  +k \eta^{abc}B{_{0a}}^{IJ}F_{bcIJ}+ k \eta^{abc}B{ _{bc}}^{IJ} \left( \dot{A}_{aIJ} - D_a A_{0IJ} \right)  \nonumber  \\  &+&  k^2\eta^{abc}B{_{0a}}^{IJ}B{_{bc}}^{IJ}.
\end{eqnarray}
We can see that  this theory  is also background independent and has the same number of dynamical variables as a pure $BF$ theory. The momenta $\left(\Pi^\alpha {_{IJ}}, \Pi^{\alpha \beta}{_{IJ}} \right)$ canonically conjugate to $\left(A^{IJ}{_{\alpha}}, B{_{\alpha \beta}}^{IJ} \right)$,   are given by 
\begin{equation}
\Pi^\alpha {_{IJ}}= \frac{\delta L}{\delta \dot{A}^{IJ}{_{\alpha}} }, \quad  \Pi^{\alpha \beta}{_{IJ}}=\frac{\delta L}{\delta \dot{B}{_{\alpha \beta}}^{IJ} }.
\label{eq29}
\end{equation}
The matrix elements of the Hessian 
\begin{equation}
\frac{\partial^2{\mathcal{L}} }{\partial( \partial_\mu (A_{\alpha} {^{IJ}} )) \partial(\partial_\mu (A_{\beta} {^{IJ}} )) },\quad \quad \frac{\partial^2{\mathcal{L}} }{\partial( \partial_\mu (A_{\alpha} {^{IJ}} ) )\partial(\partial_\mu (B{_{\rho \nu}}^{IJ} ) ) },  \quad \quad \frac{\partial^2{\mathcal{L}} }{\partial( \partial_\mu (B{_{\rho \nu}}^{IJ} )) \partial(\partial_\mu (B{_{\gamma \sigma}}^{IJ}) ) }, 
\label{eq47a}
\end{equation}
are identically zero, the rank of the Hessian is zero, thus, we expect 60 primary constraints. \\
From the definition of the momenta (\ref{eq29}),  we   identify the following  60 primary constraints
\begin{eqnarray}
\phi^0{_{IJ}} &:&   \Pi^0 {_{IJ}} \approx 0, \nonumber \\
\phi^a{_{IJ}} &:& \Pi^a {_{IJ}} -  \eta^{abc}  \left( F_{bcIJ}+ kB_{bc IJ}  \right) \approx 0,  \nonumber \\
\phi^{0a}{_{{IJ}}} &:& \Pi^{0 a}{_{IJ}} \approx 0, \nonumber \\
\phi^{ab}{_{{IJ}}} &:&\Pi^{ab}{_{IJ}} \approx  0.
\label{eq31}
\end{eqnarray}
We can appreciate that,  with respect to  the primary constraints for  a pure $BF$ theory, the primary constraints (\ref{eq31}) present an extra term  $(F_{bcIJ})$ because of the presence of the second Chern class.\\ 
By using the definition of the momenta (\ref{eq29}), we find that the canonical Hamiltonian takes the form 
\begin{equation}
H_{c}= - \int  dx^3 \left[A^{IJ}{_{0}} D_a \Pi^a {_{IJ}} + kB{_{0a}}^{IJ} \Pi^{a}{_{IJ}} \right], 
\label{}
\end{equation}
 the canonical Hamiltonian and the addition of  primary constraints allow us to identify the primary   Hamiltonian 
\begin{equation}
H_P= H_c + \int dx^3 \left[  \lambda^{IJ}{_{0}} \phi{_{IJ}}^{0} + \lambda^{IJ}{_{a}} \phi{_{IJ}}^{a}+\lambda_{0a}{^{IJ}}\phi^{0a}{_{IJ}} +\lambda_{ab}{^{IJ}}\phi^{ab}{_{IJ}}  \right].
\label{} 
\end{equation}
The non-vanishing fundamental Poisson brackets for the theory under study are given by 
\begin{eqnarray}
\{A{^{IJ}}_\mu(x^0, x), \Pi{^{\nu}}_{KL} (x^0, y) \} &=& \delta{^{\nu}}_{\mu} \frac{1}{2} \left( \delta{^{I}}_K \delta{^{J}}_L- \delta{^{I}}_L \delta{^{J}}_K  \right) \delta^3(x-y), \nonumber \\ 
\{B_{\alpha \beta}{^{IJ}}(x^0, x), \Pi{^{\mu \nu}}_{KL}(x^0,y) \} &=& \frac{1}{4} \left( \delta{^{\mu}}_{\alpha} \delta{^{\nu}}_\beta-\delta{^{\mu}}_\beta \delta{^{\nu}}_\alpha  \right) \left( \delta{^{I}}_{K} \delta{^{J}}_L -\delta{^{I}}_{L} \delta{^{J}}_K \right) \delta^3(x-y).
\label{eq34}
\end{eqnarray}
Just as in the above section, we need to identify  if the theory presents   secondary constraints. For this aim, we compute the  60 $\times$ 60 matrix whose entries are the Poisson brackets among the primary constraints (\ref{eq31}), the nonzero brackets are given by  
\begin{eqnarray}
\{ \phi^{a}{_{IJ}} (x), \phi^{bc} {_{KL}}(y) \}&=& \frac{k}{2}  \eta^{abc} (\eta_{IL}\eta_{JK}-\eta_{IK} \eta_{JL}   ) \delta^3(x,y),    
\label{eq35}
\end{eqnarray}
this matrix has rank= 36 and 24  linearly independent null-vectors. The null vectors and   consistency   conditions  imply  24 secondary constraints.  From the temporal evolution of the constraints (\ref{eq31}) and the 24 null vectors arise  the following  24 secondary constraints 
\begin{eqnarray}
\dot{\phi}^{0}{_{IJ}}&=& \{\phi^{0}{_{IJ}} (x), {H}_{P} \} \approx 0 \quad \Rightarrow \quad \psi_{IJ}:= D_a \Pi^a{_{IJ}} \approx 0. \nonumber \\
\dot{\phi}^{0a}{_{IJ}}&= &\{\phi^{0a}{_{IJ}} (x), {H}_{P} \} \approx 0 \quad \Rightarrow \quad \psi{^{0a}}_{IJ}:= k \Pi{^{a}}_{IJ} \approx 0, 
\label{eq36}
\end{eqnarray}
and the rank fix the following 36 Lagrange multipliers 
\begin{eqnarray}
\dot{\phi}^a{_{IJ}}= \{\phi^a{_{IJ}} (x), {H}_{P} \} \approx 0 \quad & \Rightarrow& \quad \left[\Pi^a{_{JL}}\eta_{KI}-\Pi^a{_{IL}}\eta_{KJ}  \right] A_0{^{KL}} - \eta^{abc}k D_b B_{0cIJ} \nonumber \\   &-&  \left[  \eta^{abc}F_{bcKJ} \eta_{IL} +\eta^{abc}F_{bcIK} \eta_{LJ} \right] A_{0}{^{KL}}  -  \eta^{abc} k \lambda_{bcIJ}\approx 0, \nonumber \\
\dot{\phi}^{ab}{_{IJ}}= \{\phi^{ab}{_{IJ}} (x), {H}_{P} \} \approx 0 \quad &\Rightarrow& \quad  \eta^{abc} \lambda_{cIJ}  \approx 0. 
\label{eq39a}
\end{eqnarray}
From consistency of  secondary constraints,  does not emerge more constraints. In this way, with all the constraints at hand we need to identify which ones correspond to first  and second class. For this aim, we compute the Poisson brackets between the primary and  secondary constraints  which are given in the following 84$\times$84  matrix whose nonzero brackets are given by  
\begin{eqnarray}
\{ \phi^{a}{_{IJ}} (x), \phi^{bc} {_{KL}}(y) \}&=& \frac{k}{2}  \eta^{abc} (\eta_{IL}\eta_{JK}-\eta_{IK} \eta_{JL}   ) \delta^3(x,y),  \nonumber \\    
\{ \phi^{a}{_{IJ}} (x), \Psi {_{KL}}(y) \}& =&\frac{1}{2}\left( \Pi^a{_{JL}} \eta_{IK} - \Pi^a {_{IL}} \eta_{KJ}  + \Pi^a {_{KJ}} \eta_{IL} - \Pi^a {_{KI}}\eta_{LJ}     \right) \delta^3(x-y), \nonumber \\
\{ \phi^{a}{_{IJ}} (x), \Psi^b {_{KL}}(y) \}& =& k \eta^{abc} \Big[ (\eta_{IK} \eta_{JL} - \eta_{KJ} \eta_{IL})\partial_c  + (A_{JLc} \eta_{IK} -A_{ILc} \eta_{KJ})\nonumber \\  & -& (A_{KI} \eta_{LJ} - A_{KJc} \eta_{LI} )\Big] \delta^3(x-y) , \nonumber \\
\{ \Psi{_{IJ}} (x), \Psi {_{KL}}(y) \}& =& - \frac{1}{2}( \Psi_{LJ} \eta _{IK} - \Psi_{KJ} \eta_{IL} + \Psi_{IL}\eta_{KJ} - \Psi_{IK}\eta_{LJ} )\delta^3(x-y) \approx 0, \nonumber \\
\{ \Psi{_{IJ}} (x), \Psi^a {_{KL}}(y) \}& =& -\frac{k}{2}\left( \Pi^a{_{JL}} \eta_{IK} - \Pi^a {_{IL}} \eta_{KJ}  + \Pi^a {_{KJ}} \eta_{IL} - \Pi^a {_{KI}}\eta_{LJ}     \right) \delta^3(x-y), \label{eq13}
\end{eqnarray}�
After long calculations, we  observe that this  matrix has a rank= 36 and 48 null-vectors.  In this manner, by using the null vectors  one finds  the following 48 first class constraints 
\begin{eqnarray}
\gamma^0{_{IJ}} &:&   \Pi^0 {_{IJ}}, \nonumber \\
\gamma^{0a}{_{IJ}} &:& \Pi^{0 a}{_{IJ}},  \nonumber \\
\gamma_{IJ}&:& D_a \Pi^{a}{_{IJ}} - \left[ \Pi{^{ab}}_{JK}B{_{abI}}^K - \Pi{^{ab}}_{IK}B{_{abJ}}^K  \right  ] ,  \nonumber \\
\gamma^a{_{IJ}} &:&k \Pi{^{a}}_{IJ} + D_b\Pi{^{ba}}_{IJ} ,
\label{eq41a}
\end{eqnarray}
where the third constraint can be identified as the Gauss constraint for this generalized $BF$ theory, corresponding  to the generator of $SO(3,1)$ transformations. Again,  we  observe that by means of the null vectors the form of the secondary constraints  has been changed and becomes to be of first class;  this fact is important because we do not need fix by  hand  the constraints to convert them in first class.  \\
 From the rank we can  identify the following  36 second class constraints
\begin{eqnarray}
\chi^a{_{IJ}} &:&\Pi^a {_{IJ}} -  \eta^{abc}  \left( F_{bcIJ}+ kB_{bc IJ}  \right),  \nonumber \\
\chi^{ab}{_{{IJ}}} &:&\Pi^{ab}{_{IJ}}. 
\label{eq42a}
\end{eqnarray}
So, we  see that does exist a clear difference among pure $BF$ theory and this generalized theory  in the set of the first and second class constraints. In fact, the corresponding $\gamma^a{_{IJ}}$ constraints and second class constraints  $\chi^a{_{IJ}} $ are quite different, however this result  is  expected due   to presence of second Chern class and the cosmological-like terms  in the action. \\ 
An other important point to observe is that the constraints (\ref{eq41a}) are not all independent because of Bianchis identity $DF=0$, that now implies
\begin{equation}
D_{a}\gamma^{a}{_{IJ}}- k\gamma{_{IJ}}- k\left[ \chi{^{ab}}_{JK}B{_{abI}}^K - \chi{^{ab}}_{IK}B{_{abJ}}^K  \right  ]-\left[ \chi{^{ab}}_{IK} F{_{abJ}}^K -  \chi{^{ab}}_{KJ} F{_{abI}}^K \right]=0,
\end{equation}
which correspond to  6 reducibility conditions for the theory. Thus, from the  $\gamma^{a}_{IJ}$=18  first class constraints   we identify that  [18-6]= 12  are  independent, just as for a pure $BF$ theory. Therefore,  we are able to  procedure and   calculate the physical degrees of freedom as follows; the theory presents  120 canonical   variables, [48-6]=42  independent first class constraints   and 36 independent second class constraints. This information allows us to  conclude that this generalized four-dimensional $BF$ theory is devoid of physical degrees of freedom and corresponds to a topological field theory as well. It is important to comment that this program     allowed  us to know the full structure of the constraints,  and now it is straightforward to know   the  results  that  can be obtained  by considering as dynamical variables those occurring with time derivative  in the action principle. \\
Now, we will calculate the algebra of  the constraints. Smearing the constraints with test fields 
\begin{eqnarray}
\phi_1 &:=& \gamma^0{_{IJ}} \left[ \mathbf{A} \right]= \int dx^3  {\mathbf{A}}^{IJ} \left[ \Pi^0{_{IJ}} \right], \nonumber \\
\phi_2  &:=&\gamma^{0a}{_{IJ}} \left[ \mathbf{B} \right]= \int dx^3 {\mathbf{B}}_{a}{^{IJ}} \left[  \Pi^{0a}{_{IJ}}  \right], \nonumber \\
\phi_3  &:=& \gamma{_{IJ}} \left[ \mathbf{C} \right]= \int dx^3 {\mathbf{C}}_a{^{IJ}}  \left[ D_a \Pi^{a}{_{IJ}} - \left[ \Pi{^{ab}}_{JK}B{_{abI}}^K - \Pi{^{ab}}_{IK}B{_{abJ}}^K  \right  ] \right], \nonumber \\
\phi_4 &:=& \gamma^{a}{_{IJ}} \left[ \mathbf{D} \right]= \int dx^3  {\mathbf{D}}{_{0a}}^{IJ} \left[  \Pi{^{a}}_{IJ} + D_b\Pi{^{ba}}_{IJ} \right], \nonumber \\
\phi_5  &:=&\chi^a{_{IJ}}  \left[ \mathbf{H} \right]= \int dx^3 {\mathbf{H}}_a{^{IJ}} \left[\Pi^a {_{IJ}} -  \eta^{abc}  \left( F_{bcIJ}+ kB_{bc IJ}  \right) \right], \nonumber \\
\phi_6  &:=&\chi^{ab}{_{IJ}} \left[ \mathbf{G} \right]= \int dx^3 {\mathbf{G}}{_{ab}}{^{IJ}} \left[ \Pi^{ab} {_{IJ}}  \right].
\label{eq44a}
\end{eqnarray} 
 the non-zero brackets  of   the constraints  are  given  by 
\begin{eqnarray}
\Big\{\phi_3\left[ {\mathbf{C}}{^{IJ}} \right],\phi_3\left[ {\mathbf{C}}'{^{KL}} \right]  \Big\} &=& \int dx^3\left[ {\mathbf{C}}^{IK} {\mathbf{C}}'_K{^{J}}-{\mathbf{C}}^{JK} {\mathbf{C}}'_K{^{I}} \right]  \gamma{_{IJ}} \approx0, \nonumber \\
\Big\{\phi_3\left[ {\mathbf{C}}{^{IJ}} \right],\phi_4\left[ {\mathbf{D}}{_{0a}}^{KL} \right]   \Big\} &=& \int dx^3\left[ {\mathbf{C}}^{IK}{\mathbf{D}}_{0aK}{^{J}}-{\mathbf{C}}^{JK}{\mathbf{D}}_{0aK}{^{I}} \right]  \gamma^{0a}{_{IJ}} \approx0, \nonumber \\
\Big\{\phi_3\left[ {\mathbf{C}}{^{IJ}} \right],\phi_5\left[ {\mathbf{H}}_a{^{KL}} \right]  \Big\} &=& \int dx^3\left[ {\mathbf{C}}^{IK} {\mathbf{H}}_{aK}{^{J}}- {\mathbf{C}}^{JK} {\mathbf{H}}{_{aK}}{^{I}} \right] \chi^a{_{IJ}}  \approx 0, \nonumber \\
\Big\{\phi_5 \left[ {\mathbf{H}}_{a}{^{IJ}}  \right] , \phi_6 \left[ {\mathbf{G}}'_{ab}{^{KL}}  \right]  \Big\}&=& -\eta^{abc}k\int dx^3 \left[ {\mathbf{H}}_{aKH} {\mathbf{G}}{_{bc}}^{KH} \right], 
\label{eq45a}
\end{eqnarray}
where we can see that the constraints (\ref{eq44a}) correspond to a set of first and second class constraints respectively. \\
The constraint algebra (\ref{eq45a}) allows us   to   identify the Dirac bracket for the theory, and    we  observe  that  the  matrix whose elements are only the Poisson brackets among   second class constraints  is given by 
\begin{equation}
C_{\alpha\beta} = \left(
\begin{array}{rr}
0 \qquad \qquad& \frac{k}{2}  \eta^{abc} (\eta_{IL}\eta_{JK}-\eta_{IK} \eta_{JL}   ) \delta^3(x,y)  \\
 -  \frac{k}{2}  \eta^{abc} (\eta_{IL}\eta_{JK}-\eta_{IK} \eta_{JL}   ) \delta^3(x,y)   & 0 \qquad \qquad \\
 \label{eqaa}
\end{array}
\right).
\end{equation}
In this manner,  the Dirac bracket among    two functionals $A$, $B$  is expressed   by 
\begin{equation}
\{A(x),B(y) \}_D= \{A(x),B(y)\}_P + \int du dv \{A(x), \zeta^\alpha(u) \} C^{-1}_{\alpha \beta}(u,v) \{\zeta^\beta(v), B(y) \}, 
\label{eq27z}
\end{equation}
where $ \{A(x),B(y)\}_P$ is the usual Poisson bracket between the functionals $A,B$,   $\zeta^\alpha(u)=(\chi{_{IJ}}^{a}, \chi{_{IJ}}^{ab} ) $ with   $C^{-1}_{\alpha \beta}(u,v)$  being   the inverse of (\ref{eqaa}) which is straightforward to obtain. We will use in future works  Dirac's bracket performing a canonical quantization scheme, since in  this paper we are only focused on a classical description of the theories under study.\\
Just as in above  section,  with the identification of the constraints as first and second class and by  using the Lagrange multipliers (\ref{eq39a}),  the constraints (\ref{eq41a}) and (\ref{eq42a}),   we find that extended action has the following form
 \begin{eqnarray}
&S_E&\left[ A_{\alpha} {^{IJ}},\Pi^{\alpha}{_{IJ}}, B{_{\mu \nu}}^{IJ}, \Pi^{\mu \nu}{_{IJ}}, u_0{^{IJ}}, u_{0a}{^{IJ}}, u{^{IJ}}, u{_{a}}^{IJ}, v_a{^{IJ}}, v{_{ab}}{^{IJ}}  \right] =\int \Big\{ \dot{A}_{\alpha} {^{IJ}}\Pi^{\alpha}{_{IJ}}+ \dot{ B}{_{0a}}^{IJ}\Pi^{0a}{_{IJ}}\nonumber \\ 
&+& \dot{ B}{_{ab}}^{IJ}\Pi^{ab}{_{IJ}} -  H-u_0{^{IJ}} \gamma^0{_{IJ}}- u_{0a}{^{IJ}} \gamma^{0a}{_{IJ}} -    u{^{IJ}} \gamma{_{IJ}}-     u{_{a}}^{IJ}\gamma^{0a}{_{IJ}}    -v_a{^{IJ}} \chi^a{_{IJ}}     -v{_{ab}}{^{IJ}} \chi^{ab}{_{IJ}}         \Big\} dx^4, \nonumber \\
\label{eq46a}
\end{eqnarray}
and here $H$ is a linear combination of first class constraints 
\begin{equation}
H= A_0{^{IJ}}\left[  D_a \Pi^{a}{_{IJ}} - \left[ \Pi{^{ab}}_{JK}B{_{abI}}^K - \Pi{^{ab}}_{IK}B{_{abJ}}^K  \right  ]  \right]- B_{0a}{^{IJ}}\left[ k \Pi{^{a}}_{IJ} + D_b\Pi{^{ba}}_{IJ} \right],
\label{eq47a}
\end{equation}
and $u_0{^{IJ}}, u_{0a}{^{IJ}}, u{^{IJ}}, u{_{a}}^{IJ}, v_a{^{IJ}}, v{_{ab}}{^{IJ}}$ are the Lagrange multipliers enforcing the constraints.\\
From (\ref{eq45a}) we similarly   identify the extended Hamiltonian  
\begin{equation}
H_E= H-u_0{^{IJ}} \gamma^0{_{IJ}}- u_{0a}{^{IJ}} \gamma^{0a}{_{IJ}} - u{^{IJ}} \gamma{_{IJ}}- u{_{a}}^{IJ}\gamma^{0a}{_{IJ}}.
\label{eq51x}
\end{equation}
Again,  it is remarkable to observe that the extended Hamiltonian (\ref{eq51x}) is a linear combination of first class constraints reflecting the general covariance of theory,  thus,  we can not construct the Shcrodinger equation because of the   action of the Hamiltonian on physical states is annihilation \cite{13}.\\
The equations of motion obtained from the extended action  are 
expressed by 
\begin{eqnarray}
\delta A{_{0}}^{IJ}:  \dot{\Pi}^0{_{IJ}}&=&- \left[  D_a \Pi^{a}{_{IJ}} - \left[ \Pi{^{ab}}_{JK}B{_{abI}}^K - \Pi{^{ab}}_{IK}B{_{abJ}}^K  \right  ]  \right] ,  \nonumber \\
\delta \Pi^0{_{IJ}}:  \dot{A}{_{0}}^{IJ} &=& u{_{0}}^{IJ},  \nonumber \\
\delta A{_{a}}^{IJ}: \dot{\Pi}^a{_{IJ}}&=&2A_{0}{^{FH}}\left[ \Pi^a{_{JH}} \eta_{FI} -\Pi^a{_{IH}}    \eta_{FJ} \right] - 2 u{^{FH}} \left[ \Pi^a{_{JH}} \eta_{FI}-\Pi^a{_{IH}}\eta_{FJ} \right] \nonumber \\ &+& 2 B_{0b}{^{FH}} \left[ \Pi^{ab}{_{JH}}\eta_{FI}- \Pi^{ab}{_{IH}}  \eta_{FJ}\right] - 2 u_b{^{FH}}\left[ \Pi^{ab}{_{JH}}\eta_{FI}- \Pi^{ab}{_{IH}}  \eta_{FJ} \right] +2 \eta^{abc} D_{c}v_{bIJ}   , \nonumber \\
\delta \Pi^a{_{IJ}}:   \dot{A}{_{a}}^{IJ}&=&-D_a\left(A_0{^{IJ}}+u^{IJ}  \right) + k \left(u_a{^{IJ}} -B_{0a}{^{IJ}} \right)
 + v_a{^{IJ}}, \nonumber \\ 
\delta B{_{0a}}^{IJ}:  \dot{\Pi}^{0a}{_{IJ}}&=& \left[ k \Pi{^{a}}_{IJ} + D_b\Pi{^{ba}}_{IJ} \right] ,  \nonumber \\
\delta \Pi^{0a}{_{IJ}}:  \dot{B}{_{0a}}^{IJ} &=&  u{_{0a}}^{IJ},  \nonumber \\
\delta B{_{ab}}^{IJ} : \dot{\Pi}^{ab}{_{IJ}}&=&- k\eta ^{abc}v_{cIJ}+  \left[A_{0I}{^{H}}  \Pi^{ab}{_{HJ}}- A_{0J}{^{H}}  \Pi^{ab}{_{HI}} \right] + \left[ u_{I}{^{H}}   \Pi^{ab}{_{HJ}}- u_{J}{^{H}}   \Pi^{ab}{_{HI}} \right], \nonumber \\
\delta \Pi^{ab}{_{IJ}}:  \dot{B}{_{ab}}^{IJ}&=&  - D_a u_b{^{IJ}} + D_b u_a{^{IJ}} + v_{ab}{^{IJ}} - \left[A_{0}{^{FI}} B_{abF}{^{J}}-A_{0}{^{FJ}} B_{abF}{^{I}} \right] , \nonumber\\  &-& \left[ u^{FI} B_{abF}{^{J}}-u^{FJ} B_{abF}{^{I}}   \right],  \nonumber \\
\delta u_0{^{IJ}} : \gamma^0{_{IJ}}&=&0, \nonumber \\
\delta  u_{0a}{^{IJ}}: \gamma^{0a}{_{IJ}}&=&0, \nonumber \\
\delta  u{^{IJ}}: \gamma{_{IJ}}&=&0, \nonumber \\
\delta u{_{a}}^{IJ}:\gamma^{0a}{_{IJ}} &=&0, \nonumber \\
\delta v_a{^{IJ}} :\chi^a{_{IJ}} &=&0, \nonumber \\
\delta v{_{ab}}{^{IJ}}: \chi^{ab}{_{IJ}}                                              &=&0 . 
\label{eq49a}
\end{eqnarray}
\newline
\newline
\noindent \textbf{V. Gauge generator }\\[1ex]
By following with our analysis, we need to know the gauge transformations of the theory. For this important step we shall use  Castellani's formalism \cite{11}, by defining  the following gauge generator in terms of the first class constraints (\ref{eq41a})
 \begin{eqnarray}
G= \int_\Sigma \left[D_0 \varepsilon_0{^{IJ}} \gamma^0{_{IJ}}  + D_0 \varepsilon_{0a}{^{IJ}} \gamma^{0a}{_{IJ}} +\varepsilon^{IJ}\gamma{_{IJ}} + \varepsilon_a {^{IJ}}\gamma^{0a}{_{IJ}}  \right]dx^3, 
\label{eq62}
\end{eqnarray}
thus, we find the following gauge transformations on the phase  space 
\begin{eqnarray}
\delta_0 A{_{0}}^{IJ} &=&D_0  \varepsilon{_{0}}^{IJ},  \nonumber \\
\delta_0 A{_{a}}^{IJ} &=& -  D_a  \varepsilon^{IJ} + k \varepsilon_{a}{^{IJ}},  \nonumber \\
\delta_0 B{_{0a}}^{IJ} &=&   D_0  \varepsilon{_{0a}}^{IJ},  \nonumber \\
\delta_0 B{_{ab}}^{IJ} &=& \left[ D_a \varepsilon{_{b}}^{IJ} - D_b\varepsilon{_{a}}^{IJ} \right] + \left[ \varepsilon^{IF} B_{abF}{^{J}}- \varepsilon^{JF}B_{abF}{^{I}} \right],  \nonumber \\
\delta_0 \Pi^0{_{IJ}}   &=&   0,  \nonumber \\
\delta_0 \Pi^a{_{IJ}}   &=& \left[\Pi^a{_{IL}}\varepsilon_J{^{L}}- \Pi^a{_{JL}}\varepsilon_I{^{L}}  \right] + \left[ \Pi^{ab}{_{KI}}\varepsilon_b {^{L}}_J- \Pi^{ab}{_{KJ}}\varepsilon_b {^{L}}_I \right],  \nonumber \\
\delta_0 \Pi^{0a}{_{I}}  &=&   0,  \nonumber \\
\delta_0 \Pi^{ab}{_{IJ}}  &=& - \left[\Pi^{ab}{_{IF}} \varepsilon^{F}{_{J}}- \Pi^{ab}{_{JF}} \varepsilon^{F}{_{I}} \right].  
\label{eq51a}
\end{eqnarray}
We have seen that the extended Hamiltonian for this theory was  linear combination of first class constraints, so  diffeomorphisms are the  gauge transformations, however   the diffeomorphisms are not present in  above transformations. Furthermore,   it is well-known  that $BF$ and Pontryagin theory are  diffeomorphisms covariant and we expect that   (\ref{eq26}) being  a coupled  theory of topological terms this important symmetry is not lost. Thus, the next question that arises is; how  can we  recover  diffeomorphisms symmetry from the former gauge transformations?. The answer can be found if we redefine now the gauge parameters as $-\varepsilon{_{0}}^{IJ}=\varepsilon^{IJ}=- \xi^\rho A_\rho{^{IJ}}$ and $\varepsilon_{\mu}^{IJ}=- \xi^\rho B_{\mu \rho}{^{IJ}}$,  in this manner  the gauge transformations (\ref{eq51a}) take the form 
\begin{eqnarray}
A'_{\mu}{^{IJ}} &\rightarrow& A_{\mu}{^{IJ}} + {\mathcal{L}}_\xi  A_{\mu}{^{IJ}} + \xi^{\rho}\left[ F_{\mu \rho}{^{IJ}} +k B_{\mu \rho}{^{IJ}}\right]  , \nonumber \\
B'_{\mu \nu}{^{IJ}}& \rightarrow& B_{\mu \nu}{^{IJ}} + {\mathcal{L}}_\xi  B_{\mu \nu}{^{IJ}} + \xi^{\rho} \left[ D_\nu B_{\mu \rho}{^{IJ}} +D_\mu B_{\rho \nu}{^{IJ}}+  D_\rho B_{\nu \mu }{^{IJ}} \right], 
\label{eq25}
\end{eqnarray}
which correspond to diffeomorphisms. Therefore, the latter are an internal symmetry of the theory.  It is important to remark,  that  this analysis   has not been reported in the literature  with the details that are displayed   here. Hence, we have performed  a complete local study for the action (\ref{eq26}) that will be useful to study  global symmetries  by using for instance,  the Atiyah-Singer theorem \cite{17}, then we can use the tools developed in  Loop Quantum Gravity to quantize this theory; we need to remember that Loop Quantum Gravity is a canonical approach  where  diffeomorphisms covariant theories are quantized without perturbative methods. \\
\newline
\newline
\newline
\section{ Conclusions and prospects}
 The Hamiltonian analysis for  four-dimensional $BF$ theories has been performed. By considering the complete  set of dynamical variables that define these   theories, we have obtained all the symmetries,   the constraints,  gauge transformations, the counting of degrees of freedom and the extended Hamiltonian. For the case of a pure $BF$ theory, the present work has extended and completed the results reported in \cite{9}, where the study was performed on a  smaller phase space context, allowing us to know the complete structure of the constraints and the algebra associated. Respect to the four-dimensional generalized $BF$ theory, and despite of there are additional terms  in the action such as the second Chern class and the cosmological-like term quadratic in  $B$ field,  we were able  to know the principal symmetries of the theory. The analysis allowed us to conclude that the  topological structure of the theory   as well as  diffeomorphisms covariance was preserved. \\
 With the present work, we have at hand a better classical description of the theories studied,  thus the approach developed along the paper is an alternative way to perform  a pure Hamiltonian framework for any theory under investigation. We can see  alternative approaches in \cite{10, 11b},  where in the former  Dirac's study for  Pleba\'nski's   and the later for Palatini theories  were performed;  however that study was not a pure Hamiltonian analysis as the present work;  for instance, in \cite{10}   alternative variables in the fields were  used to carry out the analysis and the final structure of the constraints was fixed by hand. In this sense,  we expect that our approach will be an good alternative way to study the symmetries of  Plebanski actions (see appendix below),  expecting to obtain a better description \cite{13}.\\ 
We would finish with  some remarks. Topological field theories juts as the theories studied here, are characterized by being devoid of
local degrees of freedom. That is, the theories are susceptible only to global degrees of freedom associated with non-trivial topologies of the manifold on which they are defined and the topologies of the gauge bundle \cite{1, 6, 17}. Hence, in this paper we have analyzed  local symmetries of the theories under study, however,  we would emphasize that our results  have been useful  to analyze the moduli space of  a four-dimmensional $BF$ theory for a general base manifold \cite{17}. In fact, in \cite{17}  the studio of  global symmetries  for  a $BF$ theory by employing as the main tool the Atiyah-Singer theorem has been performed  with  base manifolds as $S^4$, $K3$, $E(n)$, $S_d$, etc., from which  the dimension of the moduli space has been calculated,   showing that there exist global degrees of freedom as  expected. Therefore,   $BF$ theory has been characterized both globall and locally providing all necessary elements to make progress in the quantization; these subjects will be reported in forthcoming works.
 \newline
\newline
\newline
\noindent \textbf{Acknowledgements}\\[1ex]
This work was supported by CONACyT  M\'exico under grant 95560. One of the authors (IRG) would like to thank the Departamento de F\'isica Matem\'atica of the Instituto de Ciencias, Universidad Aut\'onoma de Puebla (B.U.A.P.) for its kind hospitality.\\
\section{ Appendix A}
In this appendix we shall develop  the standard way to carry out the canonical analysis to  Pleba\'nski theory.  Pleba\'nski's  theory is given  \cite{19}
 \begin{equation}
 S[B, A, \Phi]= \frac{2i}{\kappa}\int \left( B_i \wedge F^i + \Phi_{ij} B^{i}\wedge B^{j}  \right) dx^4, 
 \label{eq68}
 \end{equation} 
where $F$ is the two form strength of a $SU(2)$ complex one-form $A=A_i t^i$, being  $t^i$ the generators of gauge group,   $B=B^it_i$ is a Lie algebra valued two-form, and $\Phi$ is a zero-form. Now if we decompose $\Phi$ into its trace and its traceless part,  namely $\phi_{ij}=\Phi^{ij} - \frac{1}{3} \phi\delta_{ij}$, with $\phi= \Phi_{ij} \delta^{ij}$ which is related with the cosmological constant \cite{15},  we get    
 \begin{equation}
 S[B, A, \phi]= \frac{2i}{\kappa}\int \left( B_i \wedge F^i + \phi_{ij} B^{i}\wedge B^{j} - \frac{1}{3} \phi B^{i} \wedge B_i \right) dx^4 
 \label{eq68}
 \end{equation}
So,  by performing the 3+1 decomposition in (\ref{eq68}) we obtain 
\begin{eqnarray}
 S[B, A, \phi]= \frac{2i}{\kappa} \int \left( \dot{A}^{i}_a (\eta^{abc}B_{ibc}) + A_{0}^{i}D_a(\eta^{abc} B_{ibc}) + \eta^{abc} F_{ab}^{i}B_{i0a}+ \eta^{abc} \phi_{ij}B^{i}_{0a}B^{j}_{bc} - \frac{\phi}{3} \eta^{abc}B^i_{0a}B_{ibc}\right), \nonumber \\
 \label{eq71}
\end{eqnarray}
where $D_a \lambda ^{i}_b=\partial_a \lambda^i_b + \varepsilon^i{_{jk}}A^j_a \lambda^k _b$ and  $\eta^{abc}=\epsilon^{0abc}$, $a,b,c= 1, 2,3$. From (\ref{eq71})  we identify the following Lagrangian density 
\begin{equation}
{\mathcal{L}}=   \frac{2i}{\kappa}\left( \dot{A}^{i}_a (\eta^{abc}B_{ibc}) + A_{0}^{i}D_a(\eta^{abc} B_{ibc}) + \eta^{abc} F_{bc}^{i}B_{i0a}+ \eta^{abc} \phi_{ij}B^{i}_{0a}B^{j}_{bc}- \frac{\phi}{3} \eta^{abc}B^i_{0a}B_{ibc} \right), 
\end{equation}
remembering  that in a smaller face space context,   we will consider as dynamical variables those that occur in the action with time derivative, so   the momenta $P^{a}_{i}$ canonically conjugate  to $A^i_a$ are given by 
\begin{equation}
P^{a}_{i}= \frac{\delta{\mathcal{L}}}{\delta A^i_a}= \frac{2i}{\kappa} (\eta^{abc}B_{ibc}), 
\label{eq72}
\end{equation}
thus, the variation respect to the fields $A_0^{i}$, $B^{i}_{0a}$ and $\phi_{ij}$ reads 
\begin{eqnarray}
D_a P^a_i&\approx& 0,  \\
\label{eq73}
\eta^{abc} F_{bc}^{i}+  \phi{^{ij}}P^a_j -\frac{\phi}{3}P^{ia}&\approx&0, \\ 
\label{eq74a}
B_{0a}^{(i}P^{aj)} -\frac{\delta^{ij}}{3} B^k_{0a}P^{a}_k&\approx&0. 
\label{eq75}
\end{eqnarray}
We observe from the later equations that it is necessary to eliminate the $B_{0a}^{i}$ and $\phi_{ij}$  variables in order to identify the structure of the constraints. The  $B_{0a}^{i}$ and $\phi_{ij}$ variables are Lagrange multipliers, however,  is not easy to identified them by using the smaller phase space approach. \\
On the other hand, it is important to mention that the action (\ref{eq71}) has a close relation with a pure $BF$ theory  and the action is not quadratic in the $B_{0i}$ variables  \cite{10, 15}, however the action worked out in \cite{10} is quadratic in  $B_{0i}$ field  and one needs   involve extra variables  to perform the hamiltonian analysis, however the  canonical analysis reported in \cite{10} becomes to be so much  complicated.   \\
Nevertheless, in our approach the canonical analysis of the first term  of the action (\ref{eq68})  has been performed,   with those results and coupling the $\phi_{ij} B^{i}\wedge B^{j}$ and $ \phi B^{i} \wedge B_i$ terms in our developments,    we will procedure  by using a pure Hamiltonian framework   to perform  the canonical analysis  of the action (\ref{eq68}).


\begin{thebibliography}{}
\setlength{\itemsep}{-.50em}
\bibitem{1} M. Montesinos and A. Perez Phys.Rev.D77:104020,2008.
\bibitem{2}  G. T. Horowitz, Commun. Math.Phys.125 (1989) 417.
\bibitem{3}  G. T. Horowitz and M. Srednicki, Commun. Math.Phys.130 (1990) 83.
\bibitem{4}  J. F. Plebanski, {\it On the separation of Einstenian substructures}, J. Math. Phys. 18 (1977) 2511.
\bibitem{5}   Derek K. Wise, {\it MacDowellÐ Mansouri Gravity and Cartan Geometry}, gr-qc/0611154;  S. W. MacDowell and F. Mansouri, Unified geometric theory of gravity and supergravity,
Phys. Rev. Lett. 38 (1977), 739Ð742. Erratum, ibid. 38 (1977), 1376;  F. Mansouri, Phys. Rev. D16, 2456 (1977).
\bibitem{6}  Freidel L and  Starodubtsev A {\it Quantum Gravity in terms of topological observables}, arXiv: hep-th/0501191.
\bibitem{7} M. Martellini, M. Zeni,  Phys. Lett. B  401 (1997) 62; hep-th/9610090.
\bibitem{8}H. Ying, Y. Ling, R. Suan, Y. Zhong, Phys.Rev. D66 (2002) 064017.
\bibitem{9} M. Mondragon, M. Montesinos, J.Math.Phys. 47 (2006) 022301.
\bibitem{10} E.Buffenoir, M.Henneaux, K.Noui, Ph.Roche, Class.Quant.Grav. 21 (2004) 5203-5220.
\bibitem{11} D. M. Gitman and I.V.Tyutin,  {\it Quantization of fields with constraints}. ( Berlin, Germany: Springer. (Springer series in nuclear and particle physics, (1990));
A. Hanson, T. Regge and C. Teitelboim. {\it Constrained Hamiltonian Systems}, (Accademia Nazionale dei Lincei, Roma, (1978));  J. Govaerts, {\it Hamiltonian quantization and constrained dynamics},Leuven, Belgium: Univ. Pr. (1991) 371 p. (Leuven notes in mathematical and theoretical physics, B4).
\bibitem{11b} P. Peldan, Class.Quant.Grav.11:1087-1132,1994. 
\bibitem{11c}  A. Escalante,  Phys. Lett B:   676, (2009), p. 105 - 111; 
A. Escalante, Int. Jour. of Theo. Phys.  48: 2486Ð2498
DOI 10.1007/s10773-009-0035-9, 2009. 
\bibitem{12a}  S. Deser, R. Jackiw and S. Templeton,  Annals of Physics 281, 409-449 (2000). \\
\bibitem{12}A. Escalante and J. Berra, {\it  A pure Dirac's method for Maxwell and Yang-Mills theories  expressed as a constrained BF theory }, submitted to Annals of Physics  (2011).
\bibitem{13} A. Escalante and J. Berra, {\it A pure Hamiltonian analysis for Plebla\'nski theory }, in preparation (2011).
\bibitem{14} J.M.F. Labastida and  , Carlos Lozano. {\it Lectures in topological quantum field theory},  hep-th/9709192. 
\bibitem{15}S. Alexandrov and  K. Krasnov, Class.Quant.Grav.26:055005,(2009).
\bibitem{17}  R. Cartas-Fuentevilla, A. Escalante and J. Berra-Montiel {\it Dimension of the moduli space  of
BF field theories}, submitted to J. of Geom.  and Phys. (2011).
\bibitem{18} J. C. Baez, Lect.Notes Phys.543:25-94,(2000).
\bibitem{19} C. Ramirez and E. Rosales,  {\it Supergroup formulation of Plebanski theory of gravity },  AIP Conf.Proc.1116:458-460,2009.


\end{thebibliography}
\end{document}